# CHEMICAL AND NUCLEAR CATALYSIS DRIVEN BY LOCALIZED ANHARMONIC VIBRATIONS

V.I. Dubinko[a*], D.V. Laptev[b]

[a] National Science Center "Kharkov Institute of Physics and Technology", Kharkov 61108, Ukraine
*vdubinko@hotmail.com
[b] B. Verkin Institute for Low Temperature Physics and Engineering, Kharkov 61103, Ukraine

In many-body nonlinear systems with sufficient anharmonicity, a special kind of lattice vibrations, namely, Localized Anharmonic Vibrations (LAV) can be excited either thermally or by external triggering, in which the amplitude of atomic oscillations greatly exceeds that of harmonic oscillations (phonons) that determine the system temperature. Coherency and persistence of LAV may have drastic effect on chemical and nuclear reaction rates due to time-periodic modulation of reaction sites. One example is a strong acceleration of *chemical reaction* rates driven by thermally-activated 'jumps' over the reaction barrier due to the time-periodic modulation of the barrier height in the LAV vicinity. At sufficiently low temperatures, the reaction rate is controlled *by quantum tunneling* through the barrier rather than by classical jumping over it. A giant increase of sub-barrier transparency was demonstrated for a parabolic potential well with the time-periodic eigenfrequency, when the modulation frequency exceeds the eigenfrequency by a factor of ~2 (parametric regime). Such regime can be realized for a hydrogen or deuterium atom in metal hydrides/deuterides, such as NiH or PdD, in the vicinity of LAV. We present an *analytical solution* of the Schrödinger equation for a *nonstationary* harmonic oscillator, analyze the parametric regime in details and discuss its applications to the tunnel effect and to D-D fusion in PdD lattice. We obtain simple analytical expressions for the increase of amplitude and energy of *zero-point oscillations* (ZPO) induced by the parametric modulation. Based on that, we demonstrate a drastic increase of the D-D fusion rate with increasing number of modulation periods evaluated in the framework of Schwinger model, which takes into account suppression of the Coulomb barrier due to lattice vibrations.

## 1. Introduction

Catalysis is at the heart of almost every chemical or nuclear transformation process, and a detailed understanding of the active species and their related reaction mechanism is of great interest [1-2]. There is no single theory of catalysis, but only a series of principles to interpret the underlying processes. An important parameter of the reaction kinetics is the *activation energy*, i.e. the energy required to overcome the reaction barrier. The lower is the activation energy, the faster the reaction rate, and so a catalyst may be thought to reduce somehow the activation energy. Dubinko et al [3-8] have shown that in a crystalline matrix, the activation energy may be reduced at some sites due to a special class of *localized anharmonic vibrations* (LAV) of atoms, known also as *discrete breathers* or *intrinsic localized modes* arising in regular crystals. LAV can be excited thermally [3, 4] or by irradiation [3, 5], resulting in a drastic acceleration of *chemical reaction* rates driven by thermally-activated 'jumps' over the reaction barrier due to the time-periodic modulation of the barrier height in the LAV vicinity. However, at sufficiently low temperatures, the reaction rate is controlled *by quantum tunneling* through the barrier rather than by classical jumping over it. The tunneling probability averaged over the Boltzmann distribution for the energy E is given by an integral [9]

$$\Gamma = \frac{1}{kT}\int_0^\infty G(E)\exp\left(-\frac{E}{kT}\right)dE \qquad (1)$$

with the tunneling coefficient (TC) given by the Gamow factor

$$G(E) \approx \exp\left\{-\frac{2}{\hbar}\int_{r_1}^{r_2}dr\sqrt{2\mu(V(r)-E)}\right\}, \qquad (2)$$

where $2\pi\hbar$ is the Planck constant, $V(r)$ is the potential barrier $\mu$ is the reduced mass, $r_1$, $r_2$ are the two classical turning points for the potential barrier.

This approach assumes that <u>dynamical behavior of the reactants does not affect the TC</u>, which is fully described by their energies. However, tunnel effect is inherently related to the operation of the *uncertainty principle* for motion along one co-ordinate, which have been generalized with account of *correlation effects* by Schrödinger [10] and Robertson [11], resulting in the following uncertainty relation (UR):

$$\sigma_x \sigma_p \geq \hbar_{ef}^2 / 4, \qquad \hbar_{ef} \equiv \hbar \times \sqrt{\left(1 + 4\frac{\sigma_{xp}^2}{\hbar^2}\right)}, \qquad (3)$$

$$\sigma_x = \left\langle\left(x-\langle x\rangle\right)^2\right\rangle, \sigma_x = \left\langle\left(p-\langle p\rangle\right)^2\right\rangle, \qquad (4)$$

$$\sigma_{xp} = \langle\hat{x}\hat{p}+\hat{p}\hat{x}\rangle/2 - \langle x\rangle\langle p\rangle, \qquad (5)$$

Dubinko V.I., Laptev D.V

where $\sigma_{xp}$ is the *mutual dispersion* between the coordinate, $x$, and momentum, $p$. At $\sigma_{xp} = 0$ (non-correlated state), and eq. (3) is reduced to the well-known Heisenberg UR, whereas at $\sigma_{xp} > 0$ one has $\hbar_{ef} > \hbar$ [1], which increases the uncertainty of coordinate and momentum and hence the tunnel probability may increase as $G^{\hbar/\hbar_{ef}}$. Therefore, the question arises about the conditions that could bring the reactant in a *coherent correlated state* (CCS) [13], which are the quantum states corresponding to the equality in the UR (3) and non-zero mutual dispersion.

Vysotskii et al [13] demonstrated that a CCS can be formed in a parabolic potential well with the time-periodic eigenfrequency. It appears that an optimal modulation frequency $\Omega$ that results in the most rapid increase of $\sigma_{xp}$ is close to $2\omega_0$ (parametric frequency): $|\Omega - 2\omega_0| \le g\omega_0$, where $\omega_0$ is the mean eigenfrequency and $g$ is the modulation amplitude.

Dubinko [7] has argued that such regime can be realized for a hydrogen or deuterium atom in metal hydrides/deuterides, such as NiH or PdD, in the vicinity of so called *gap breathers* – a sub-class of LAV arising in the H/D sub-lattice. A large mass difference between the metal and H/D atoms provides a wide phonon gap, in which gap breathers exist. Based on the numerical calculations of $\hbar_{ef}$ obtained in [13], it has been shown that the tunneling probability for the D-D fusion under electrolysis in heavy water may increase enormously with increasing number of LAV cycles resulting in the fusion rates comparable with experimental data.

In the present paper, we present an analytical solution of the Schrödinger equation for a *nonstationary* harmonic oscillator, analyze the parametric regime $\Omega = 2\omega_0$ in details and discuss its applications to the tunnel effect.

## 2. Solution of the Schrödinger equation for a nonstationary harmonic oscillator

Consider a harmonic oscillator with time-dependent frequency for a particle with the mass $m$ obeying the nonstationary Schrödinger equation of the form

$$i\hbar \frac{\partial \psi}{\partial t} = -\frac{\hbar^2}{2m} \frac{\partial^2 \psi}{\partial x^2} + \frac{m\omega^2(t)}{2} x^2 \psi. \quad (6)$$

The solution of the equation (6) can be expressed using the Green's function (or propagator):

$$\psi(x,t) = \int_{-\infty}^{+\infty} dx_0 G(x,t;x_0,t_0) \psi(x_0,t_0) \quad (7)$$

The propagator $G(x,t;x_0,t_0)$ satisfies the Schrödinger equation (6) and the following initial condition:

---
[1] Note that the definition of $\hbar_{ef}$ by eq. (3) is more straightforward than the one used in refs [12, 13], but both definitions are mathematically equivalent.

$$\lim_{t \to \tau+0} G(x,t;x_0,t_0) = \delta(x - x_0) \quad (8)$$

The expression for the propagator has the form [14]:

$$G(x,t;x_0,t_0) = \sqrt{\frac{m}{2\pi i \hbar Z}} \exp(\theta_G), \quad (9)$$

$$\theta_G = \frac{im}{2\hbar Z}\left[\frac{dZ}{dt}x^2 - 2xx_0 + Yx_0^2\right], \quad (10)$$

where the functions $Y = Y(t)$, $Z = Z(t)$ are defined by the following equations and initial conditions that can be derived from the condition (8):

$$\frac{d^2 Y}{dt^2} + \omega^2(t) Y = 0, \quad (11)$$

$$\frac{dY(t_0)}{dt} = 0, \qquad Y(t_0) = 1 \quad (12)$$

$$\frac{d^2 Z}{dt^2} + \omega^2(t) Z = 0, \quad (13)$$

$$\frac{dZ(t_0)}{dt} = 1, \qquad Z(t_0) = 0 \quad (14)$$

The functions $Y(t), Z(t)$ satisfy the condition [14]:

$$Y(t)\frac{dZ(t)}{dt} - Z(t)\frac{dY(t)}{dt} = 1. \quad (15)$$

Consider the initial wave function of the Gaussian form [15]:

$$\psi(x_0, t_0 = 0) = \frac{1}{\sqrt[4]{\pi \xi^2}} \exp\left(-\frac{x_0^2}{2\xi^2}\right), \quad (16)$$

where the characteristic length is given by

$$\xi = \sqrt{\hbar/m\omega_0}. \quad (17)$$

Then the expression for the wave function for the arbitrary moment of time $\forall t > t_0 = 0$ can be obtained from equations (7), (9), (10), (16):

$$\psi(x,t) = \frac{1}{\sqrt[4]{\pi \xi^2}} \frac{\exp(\theta_\psi)}{\sqrt{Y + i\omega_0 Z}}, \quad (18)$$

$$\theta_\psi = -\frac{x^2}{2\xi^2} \frac{1}{i\omega_0 Z}\left[\frac{dZ}{dt} - \frac{1}{Y + i\omega_0 Z}\right] \quad (19)$$

The probability density of finding the particle at $(x, t)$ is given by the square of the wave function, while the x, p and x-p dispersions are given by the following expressions:

Dubinko V.I., Laptev D.V

$$|\psi(x,t)|^2 = \frac{B(t)}{\xi\sqrt{\pi}}\exp\left\{-\frac{x^2}{\xi^2}B^2(t)\right\}, \quad (20)$$

$$B(t) = \frac{1}{\sqrt{Y^2(t)+\omega_0^2 Z^2(t)}}. \quad (21)$$

$$\langle \hat{x}\rangle = 0, \quad \langle \hat{p}\rangle = 0 \quad (22)$$

$$\sigma_x(t) = \frac{\hbar}{2m\omega_0}\left[Y^2 + \omega_0^2 Z^2\right], \quad (23)$$

$$\sigma_p(t) = \frac{\hbar m\omega_0}{2}\left[\left(\frac{1}{\omega_0}\frac{dY}{dt}\right)^2 + \left(\frac{dZ}{dt}\right)^2\right], \quad (24)$$

$$\sigma_{xp}(t) = \frac{\hbar}{2}\left[\frac{Y}{\omega_0}\frac{dY}{dt} + \omega_0 Z \frac{dZ}{dt}\right], \quad (25)$$

Consider special cases of interest.

*2.1. Constant eigenfrequency:*

$$\omega(t) = \omega_0 = const \quad (26)$$

The wave function (18) for $t > t_0 = 0$ takes the form:

$$\psi(x,t) = \frac{1}{\sqrt[4]{\pi\xi^2}}\exp\left\{-\frac{x^2}{2\xi^2} - \frac{i\omega_0 t}{2}\right\}, \quad (27)$$

whence it follows that the x and p dispersions are constant

$$\sigma_x = \frac{\hbar}{2m\omega_0}, \qquad \sigma_p = \frac{\hbar m\omega_0}{2}, \quad (28)$$

as well as the mean kinetic, potential and total energy:

$$\langle E_k\rangle = \langle E_p\rangle = \frac{\langle E\rangle}{2} = \frac{\hbar\omega_0}{4}, \quad (29)$$

while the mutual x-p dispersion is zero: $\sigma_{xp} = 0$

*2.2. Time-periodic eigenfrequency*

Consider the Mathieu equation [16] that has the same form as eqs. (11) and (13):

$$\ddot{x} + \omega_0^2\left[1 - g\cos(2\omega_0 t)\right]x = 0, \quad (30)$$

which solution can be written explicitly in the first approximation to the <u>small modulation amplitude</u> $g \ll 1$:

$$x(t) = a(t)\cos\left[\omega_0 t + \vartheta(t)\right], \quad (31)$$

where

$$a(t) = \sqrt{u^2(t) + v^2(t)}, \quad (32)$$

$$\vartheta(t) = \arctan\frac{v(t)}{u(t)}, \quad (33)$$

$$u(t) = C_1 e^{\eta} + C_2 e^{-\eta}, \quad (34)$$

$$v(t) = -C_1 e^{\eta} + C_2 e^{-\eta}, \quad \eta = \frac{g\omega_0 t}{4}. \quad (35)$$

Then the approximate solutions of the Cauchy problems (11)-(14) are given by:

$$Z(t) = \frac{1}{\omega_0}\sinh(\eta)\cos(\omega_0 t) + \\ + \frac{1}{\omega_0}\cosh(\eta)\sin(\omega_0 t) + O(g) \quad (36)$$

$$Y(t) = \cosh(\eta)\cos(\omega_0 t) + \\ + \sinh(\eta)\sin(\omega_0 t) + O(g) \quad (37)$$

Substituting eqs. (36), (37) in (23) - (25) one obtains the first approximations for dispersion of the coordinate and momentum:

$$\sigma_x(t) = \frac{\hbar}{2m\omega_0}\cosh\left(\frac{g\omega_0 t}{2}\right)\times \\ \times\left[1 + \tanh\left(\frac{g\omega_0 t}{2}\right)\sin(2\omega_0 t)\right] \quad (38)$$

$$\sigma_p(t) = \frac{\hbar m\omega_0}{2}\cosh\left(\frac{g\omega_0 t}{2}\right)\times \\ \times\left[1 - \tanh\left(\frac{g\omega_0 t}{2}\right)\sin(2\omega_0 t)\right] \quad (39)$$

The first approximation of the mutual x-p dispersion is given by

$$\sigma_{xp}(t) = \frac{\hbar}{2}\sinh\left(\frac{g\omega_0 t}{2}\right)\cos(2\omega_0 t) \quad (40)$$

Finally, the first approximation for the mean energy takes a simple form:

$$\langle E\rangle = \frac{1}{2m}\sigma_p + \frac{m\omega^2(t)}{2}\sigma_x = \\ = \frac{\hbar\omega_0}{2}\cosh\frac{g\omega_0 t}{2} \quad (41)$$

The most evident result of the parametric modulation of a parabolic potential well is increase of the coordinate, momentum and mutual dispersion with increasing number of oscillation periods, *N*, which results in rapidly increasing probability to find the oscillating particle far beyond the characteristic length of the stationary well $\xi$ (Fig. 1). It means that the amplitude of the oscillating factor $\hbar_{ef}/\hbar$ grows with *N*, but the most intriguing <u>new result</u> is a rapid growth of the oscillator *zero-point oscillation (ZPO)* energy (eq. (41) and ZPO amplitude in x and p space, which deserves a special attention as argued bellow.

## 3. Zero-point oscillation amplification

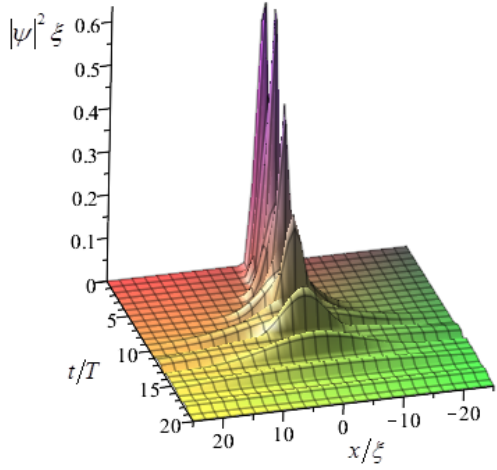

**Fig. 1**. Localization probability distribution vs. the number of oscillation periods $N = \omega_0 t / 2\pi = t/T$ in the parametric regime $\Omega = 2\omega_0$ at $g = 0.1$.

Continuous ZPO energy increase (Fig. 2) is different from the quantum energy increase to the higher oscillation levels: $E_n = \hbar\omega_0 (1/2 + n)$, when the probability density becomes concentrated at the classical "turning points". In contrast to that, we clearly deal with the ground (zero-point) state, in which the probability density is concentrated at the origin, which means the particle spends most of its time at the bottom of the potential well. However, the dispersion of its position and momentum increases along with its zero-point energy due to the parametric modulation. It is well known that zero-point energy can be derived from the uncertainty principle [17], and it is determined only by the eigenfrequency and the Plank constant: $E_0 = \hbar\omega_0 / 2$. Let us define a *ZPO amplification factor* as the ratio of the zero-point energy (eq. (41)) to its stationary value:

$$A_N = \frac{E_N}{E_0} = \cosh\frac{g\omega_0 t}{2} = \cosh(g\pi N), \quad (42)$$

which is shown in Fig. 2 along with the oscillating amplification factors for kinetic and potential energies. In contrast to the latter, ZPO amplification factor grows adiabatically with time. It is known that tunnel effect is inherently related to the operation of the uncertainty principle similar to the ZPO energy, the difference being that for the tunnel effect the coordinate is one in which the potential energy passes through a maximum, whereas for ZPO energy it passes through a minimum [9].

By equating the average potential energy, $m\omega_0^2 \Lambda^2 / 2$ to half the ZPO energy $E_N / 2$ one obtains the mean square displacement from the equilibrium position

$$\Lambda_N^2 = \frac{E_N}{m\omega_0^2} = A_N \frac{\hbar}{2m\omega_0} = A_N \Lambda_0^2, \quad \Lambda_0 = \sqrt{\frac{\hbar}{2m\omega_0}}, \quad (43)$$

where $\Lambda_0$ is the ZPO amplitude in a stationary state.

Note that the amplitude of oscillating factor $\hbar_{ef}/\hbar$ deduced from the Schrödinger-Robertson UR (eq. (3)) with account of eq. (40) coincides with amplitude of amplification of kinetic and potential energy shown in Fig. 2. In the following section, Eqs. (42)-(43) will be used for the evaluation of the D-D fusion rate in the PdD lattice.

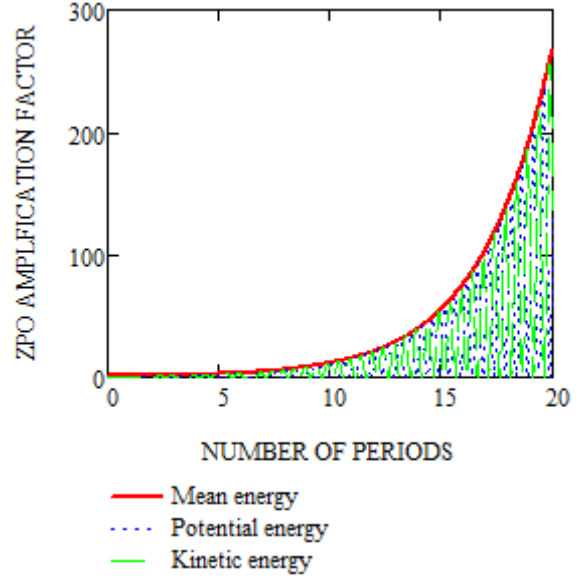

**Fig. 2**. Ratio of the zero-point energy to its stationary value in the parametric regime at $g = 0.1$ according to eq. (42).

## 4. D-D fusion in PdD lattice

According to Parmenter and Lamb (P&L) [18], the total effective potential for a deuteron pair in PdD lattice can be described by the Thomas-Fermi method, and it is given by the sum of two terms:

$$V_{eff}(r) \approx \frac{m\omega_0^2}{2} r^2 + \frac{e^2}{R_0 - r} \exp\left(-\frac{R_0 - r}{\lambda_D}\right), \quad (44)$$

where the first term is the harmonic potential well (HPW) formed by conduction electrons, in which a deuteron is trapped, and the second term is the Coulomb repulsion between the deuterons screened by the electrons; $r$ is the displacement from the equilibrium position, $R_0$ is the D-D equilibrium distance, $m$ is their mass, $e$ is the electron charge and $\lambda_D$ is the Debye screening length. At $(R_0 - r) \to r_{nucl} \sim 3 \times 10^{-5}$ Å, the barrier $V_{eff} \to 0.44$ MeV is very high but finite and narrow.

We will take the following potential parameters: the eigenfrequency $\omega_0 = 50$ THz is based on the neutron scattering analysis of DOS in PdD$_{0.63}$ crystal by Rowe at al [19] used for the gap breather analysis in [7]. The screening length $\lambda_D = 0.046$ Å corresponds to the screening potential of 310 eV measured by the yields of protons or neutrons

Dubinko V.I., Laptev D.V

emitted in the D(d, p)T or D(d, n)3He reactions induced by bombardment of D-implanted Pd [20].

In addition to the electron screening considered by P&L, a substantial suppression of the Coulomb barrier may be possible at the expense of lattice vibrations, as was argued by the Nobel Laureate Julian Schwinger [21, 22], who was the first to point out at the <u>bridge between the lattice vibrations and nuclear fusion</u>. According to [21] the effective potential of the d+d and p+d interactions is modified due to averaging ${}_0\langle\ \rangle_0$ related to their ZPO in adjacent harmonic potential wells, where ${}_0\langle\ \rangle_0$ symbolizes the phonon vacuum state. The resulting effective interaction potential is given by eq. (29) in [21]:

$$
{}_0\langle V_c(r)\rangle_0 = \frac{e^2}{r}\sqrt{\frac{2}{\pi}}\int_0^{r/\Lambda_0} dx\exp\left(-\tfrac{1}{2}x^2\right) \approx
$$

$$
\begin{cases} r \gg \Lambda_0 : \dfrac{e^2}{r} \\ r \ll \Lambda_0 : \left(\dfrac{2}{\pi}\right)^{1/2}\dfrac{e^2}{\Lambda_0} \end{cases} \qquad (45)
$$

Accordingly, the rate of fusion has been evaluated by Schwinger [22] as the rate of transition out of the phonon vacuum state, which is reciprocal of the mean lifetime $T_0$ of the vacuum state:

$$
\frac{1}{T_0} = (2\pi/\hbar)_0\langle V\delta(H-E)V\rangle_0,\ H = \frac{m\omega_0^2}{2}r^2 + V \quad (46)
$$

where $H$ is the system Hamiltonian, $E$ the energy, and $V$ is the anharmonic addition to the potential energy.

After a lengthy math, Schwinger derives a surprisingly simple estimation for the fusion rate given by

$$
\frac{1}{T_0} \sim \left(\frac{r_{nucl}}{\Lambda_0}\right)^3 \exp\left[-\frac{1}{2}\left(\frac{R_0}{\Lambda_0}\right)^2\right] \sim 10^{-19}\ s^{-1}, \quad (47)
$$

at $r_{nucl} = 10^{-5}$ Å and $\Lambda_0 = 0.1$ Å; $R_0 = 0.94$ Å deduced from X-ray measurements on hydrided Pd, and $R_0$ was the equilibrium spacing of two deuterons placed in *one site* in a hypothetical PdD$_2$ lattice. Even at such separation, the resulting fusion rate was too low to explain the observed excess heat generated in Pd cathode under D$_2$O electrolysis.

Now consider evolution of the localization probability distribution in the HPW in parametric modulation regime with increasing N shown in Fig. 3. The ZPO amplitude increases with N up to 2.5 Å at N=17 (for $\omega_0 = 50$ THz) or N=25 (for $\omega_0 = 320$ THz), as shown in Fig. 4, which defines the validity domain of the HPW approximation, beyond which the total potential at the classical turning point deviates strongly from HPW. Note that the validity domain shifts upward with increasing N with account of Schwinger effect, while the maximum effective barrier height decreases, as shown in Fig. 5.

At N=17, ZPO energy reach a level of several eV (Fig. 6), which in itself is too low for any significant tunneling through the Coulomb barrier. However, taking into account the ZPO effect by Schwinger and ZPO amplification factor given by eq. (42), the fusion rate given by eq. (47) increases drastically with increasing N:

$$
\frac{1}{T_N} \sim \left(\frac{r_{nucl}}{\Lambda_N}\right)^3 \exp\left[-\frac{1}{2}\left(\frac{R_0}{\Lambda_N}\right)^2\right], \quad (48)
$$

which is illustrated in Fig. 7.

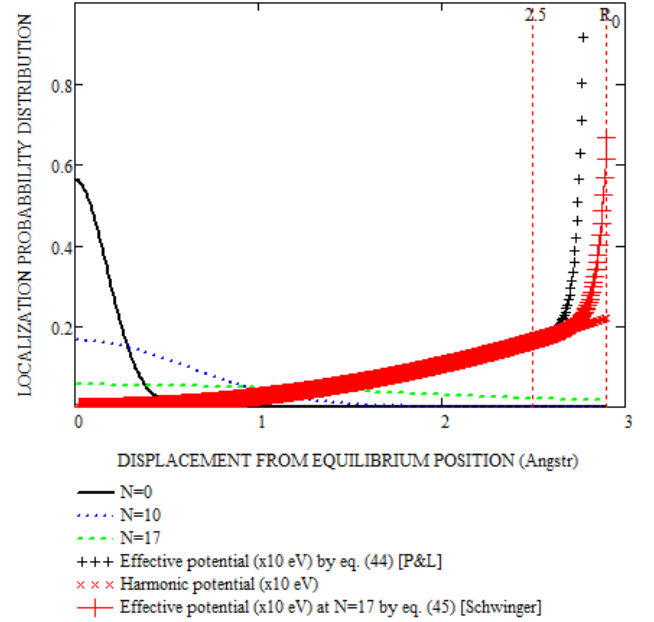

**Fig.3.** Localization probability distribution in the HPW shown by red (**x**) at different N in the parametric regime $\Omega = 2\omega_0 = 100$ THz, $g = 0.1$.

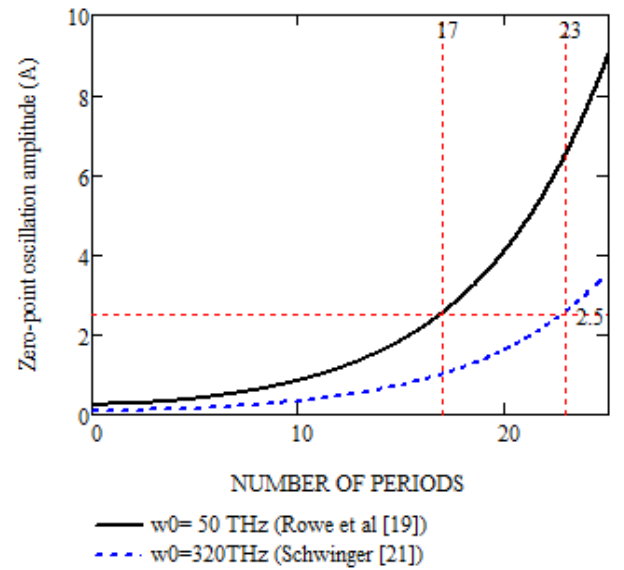

**Fig.4.** Zero-point energy increase in the HPW in the parametric regime $\Omega = 2\omega_0 = 100$, THz, $g = 0.1$.

## 5. Discussion

The parametric modulation of the HPW analyzed in the present paper was suggested to take place for a hydrogen or

Dubinko V.I., Laptev D.V

deuterium atoms in metal hydrides/deuterides, such as NiH or PdD, in the vicinity of *gap breathers* in a regular lattice [7] or LAV arising in small clusters [4, 8]. Their existence and stability is of *nonlinear origin*, which poses an important question of the nonlinearity impact on the correlation effects found for a modulated HPW.

Another important problem concerns the practical ways of LAV excitation in crystals and clusters. Heating helps to excite LAV since it enhances thermal fluctuations of atoms from equilibrium positions [3], but the LAV lifetime (which determines the number of periods in the parametric regime) is expected to decrease with increasing temperature and hence, the catalytic efficiency of LAV may decrease drastically. Therefore, we need ways to excite LAV at sufficiently low temperatures, which can be done by applying gamma, electron or ion irradiation in the energy range suitable for displacement of H/D atoms sufficiently far away from their equilibrium positions to enter nonlinear vibration regime but not too far, in order to avoid formation of structural defects.

The work by Chernov et al [23] on the excitation of hydrogen subsystems in metals by external influence give a strong support to this view. They conclude that '*under external energy input (for instance by means of radiation)* <u>*an excitation of vibrations occurs in the hydrogen subsystem*</u>*. The following facts point to this: intensive migration, diffusion and release of hydrogen isotopes from metals at low temperature; super-linear dependence of H, D release from metals on the electron current density and H, D concentration; H and D release from the whole volume of samples during the irradiation process by focused electron beam; H and D release in both molecular and atomic forms*'.

Note that all the listed phenomena belong to the realm of chemical reactions, which accompany the 'excess heat' and nuclear products measured in these experiments. It shows that both nuclear and chemical reaction triggered by 'external influence' have the same origin, and LAV is a good candidate to be the one.

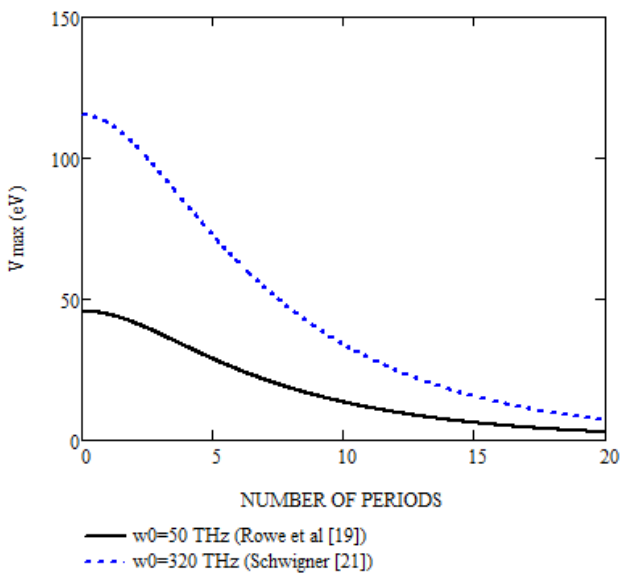

**Fig.5.** The maximum effective barrier height with account of ZPO effect by Schwinger.

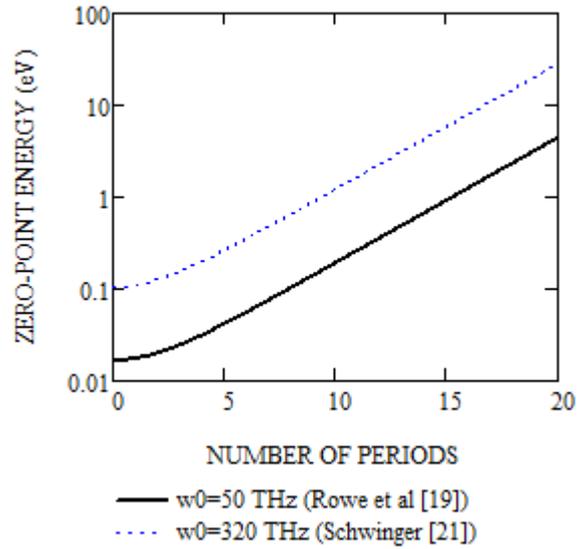

**Fig.6.** Zero-point energy increase in the HPW in the parametric regime at $g = 0.1$.

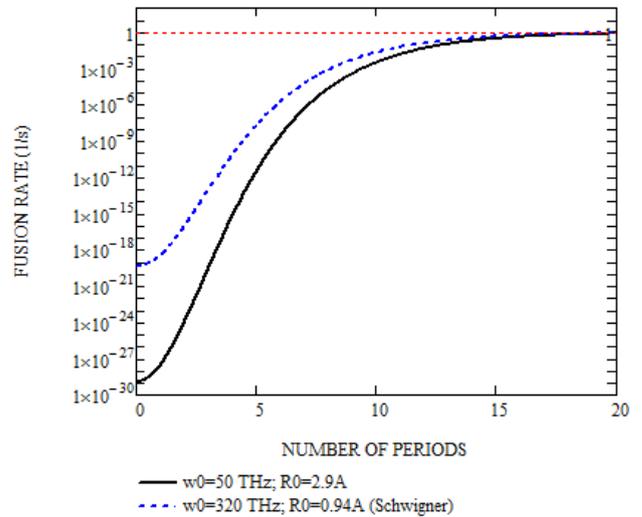

**Fig.7.** D-D fusion rate vs. N, according to eq. (48) with account of eqs. (42) and (43).

## 6. Conclusions and outlook

Analytical solution of the Schrödinger equation for a periodically driven harmonic oscillator is derived.
The oscillator zero-point energy, which is inherently related to the operation of the uncertainty principle, is shown to increase in response to parametric modulation. Based on that, a drastic increase of the D-D fusion rate with increasing number of modulation periods was demonstrated in the framework of Schwinger model, which takes into account suppression of the Coulomb barrier due to lattice vibrations.

The present concept may provide a basis for the *low energy* nuclear reactions in solids as well as for the *low temperature* chemical reactions controlled by the tunnel effect.

Atomistic modeling of LAV of various types in metal hydrides/deuterides is an important outstanding problem since it may offer the ways of *engineering* the chemical and nuclear catalysts.

Dubinko V.I., Laptev D.V

**Acknowledgements**

The author is grateful to George Chechin and Vladimir Vysotskii for helpful discussions and valuable criticism.